\documentstyle{elsart}
\begin{document}

\def \as{{\alpha_{\rm s}}}
\def \rv{{\bf r}}
\def \xv{{\bf x}}
\def \yv{{\bf y}}
\def \zv{{\bf z}}

\begin{frontmatter}
\rightline{IFUM 537/FT}
\vskip 1truecm
\title{Nonperturbative evaluation of a field correlator\\
appearing in the heavy quarkonium system}

\author {M. Baker}
\address{University of Washington, Seattle, WA 98105, USA}
\author {J. S. Ball}
\address{University of Utah, Salt Lake City, UT 84112, USA}
\author {N. Brambilla}
\address{Dipartimento di Fisica, Universit\`a di Milano, \\
INFN, Sezione di Milano, Via Celoria 16, 20133 Milano, Italy}
\author {A. Vairo}
\address {Dipartimento di Fisica, Universit\`a di Bologna, \\
Via Irnerio 46, 40126 Bologna, Italy}

\begin{abstract}
Recently an electric and magnetic field correlator appearing 
in the  description of the heavy quarkonium system was evaluated
on the lattice.  Here, we give a nonperturbative
analytical evaluation of this field correlator using a dual description
of long distance Yang--Mills theory  
and using the stochastic vacuum model. The two predictions are
 both compatible with lattice data  but show  
a different dependence on the quark separation. 
We discuss the analytic results in relation to the lattice data.
\end{abstract}

\end{frontmatter}

\newpage        

\pagenumbering{arabic}

\section{\bf Introduction}

The semirelativistic interaction for a heavy quark-antiquark 
system is known in terms of electric and magnetic field correlators 
\cite{Wilson,NRQCD,DoSi,BV,BCP,BMP,DQCDmal}. Due to confinement, these 
correlators cannot simply be evaluated in perturbation theory. However, 
it is possible to obtain them with lattice simulations. Analytic 
estimates of the correlator behaviour can be obtained by making some 
assumptions on the nature of the QCD vacuum or on the confinement mechanism. 
Until now the lattice results are compatible with, 
up to the available precision, 
the analytic calculations of the field correlators involved in the static 
and spin-dependent quark-antiquark semirelativistic interaction 
(see e.g. \cite{BV}). Recently, the correlators 
entering the velocity--dependent 
spin-independent part of the interaction were obtained on 
the lattice \cite{Bali} and are compatible with the existing predictions 
(see \cite{DoSi,BCP,BMP,DQCDmal}). 

In \cite{BMP} it was shown that a spin independent term, 
${\displaystyle{{1\over 8}\left({1\over m_1^2} + {1\over m_2^2}\right)}} 
\Delta V_{\rm a}$, arises at order $1/m^2$ in the quark-antiquark 
potential from the Darwin and the spin-spin interaction. 
A previous analytic  determination of $\Delta V_{\rm a}$
in the minimal area law model   \cite{BMP} gives zero.
The first lattice QCD evaluation of 
 $\Delta V_{\rm a}$ has just been carried out and gives non--zero
negative value  \cite{Bali}. 
For quark-antiquark distances greater than 0.2 $\sim$ 0.3 fm 
its behaviour is compatible with a negative constant with a value between 
0 and $- 1.5$ GeV$^3$. For distances shorter than 
0.2 fm $\Delta V_{\rm a}$ falls off like $1/r$.

In this letter we calculate $\Delta V_{\rm a}$ 
using a dual description of long distance Yang--Mills theory
 \cite{DQCD}. 
We give also the predictions of the stochastic vacuum model \cite{DoSi}
thereby completing  a recently given comparison between these 
models \cite{BV}. Finally we compare the theoretical 
predictions with the lattice measurements and discuss 
the implication of these results for the confinement mechanism.

$\Delta V_{\rm a}$ is defined as following
\footnote{
Here and in the following $e$ stands for the usual QCD coupling constant, 
$\as\equiv e^2/4 \pi$, while $g$ denotes the dual coupling 
constant $ g\equiv 2 \pi / e$. We have used this unconventional notation 
to make contact with the dual theory papers.}: 
\begin{eqnarray}
\Delta V_{\rm a} &=& \Delta V_{\rm a}^{B} + \Delta V_{\rm a}^{E}  \> ,
\label{va}\\
\Delta V_{\rm a}^B &=& 
- \lim_{T\to\infty} {ie^2\over T} \int_{-T/2}^{T/2} dt 
\int_{-T/2}^{T/2}  dt^\prime   
\langle\!\langle {\bf B}(z_1)\cdot {\bf B}(z_1^\prime)\rangle\!\rangle  
- \langle\!\langle {\bf B}(z_1) \rangle\!\rangle \cdot 
\langle\!\langle {\bf B}(z_1^\prime)\rangle\!\rangle  \>, \nonumber\\
\label{vab}\\
\Delta V_{\rm a}^E &=& 
\lim_{T\to\infty} {ie^2\over T} \int_{-T/2}^{T/2} dt 
\int_{-T/2}^{T/2}  dt^\prime   
\langle\!\langle {\bf E}(z_1) \cdot {\bf E}(z_1^\prime)\rangle\!\rangle  
- \langle\!\langle {\bf E}(z_1) \rangle\!\rangle \cdot 
\langle\!\langle {\bf E}(z_1^\prime)\rangle\!\rangle  \>, 
\nonumber\\ \label{vae}
\end{eqnarray}
where
\begin{eqnarray}
E_i = F_{0i} \, , \quad\quad  B_i = \hat{F}_{0i} \, , \quad\quad 
F_{\mu\nu} &=& \partial_\mu A_\nu - \partial_\nu A_\mu 
+ i e \, [A_\mu, A_\nu] \, , \nonumber\\
\hat{F}^{\mu \nu} &=& {1\over 2} \epsilon^{\mu \nu \rho \sigma}  
F_{\rho\sigma} \, , 
\nonumber
\end{eqnarray}
and 
\begin{equation}
\langle\!\langle f(A) \rangle\!\rangle =
{\displaystyle{\int {\cal D} A \, e^{iS_{\rm YM} (A)} 
{\rm Tr \>}{\rm P \>} \left\{ f(A) 
\exp \left[i e \oint_{\Gamma} dx^\mu A_\mu (x) \right] \right\} }
\over \displaystyle{
\int {\cal D} A\,  e^{iS_{\rm YM} (A)} {\rm Tr \>}{\rm P \>}
\exp \left[i e \oint_{\Gamma} dx^\mu A_\mu (x) \right] } } \, .
\label{twobra}
\end{equation}
We define, also, the Wilson loop:
\begin{equation}
W(\Gamma) =  
{\rm P\>} \exp \left[i e \oint_{\Gamma} dx^\mu A_\mu (x) \right] \> . 
\label{wgamma}
\end{equation}
The closed loop $\Gamma$ is defined by the quark (anti-quark) 
trajectories ${\bf z}_1 (t)$ (${\bf z}_2 (t)$) where $t$ varies from the 
initial time $-T/2$ to the final time $T/2$.
The quark (anti-quark) trajectories ${\bf z}_1(t)$  (${\bf z}_2 (t)$) 
define the world lines $\Gamma_1$ ($\Gamma_2$). 
The world lines $\Gamma_1$ and $\Gamma_2$, along with two straight-lines 
at fixed time connecting the end points,  make up the contour $\Gamma$. 
We use also the notation $z_1 \equiv z_1(t)$ and 
$z_1^\prime \equiv z_1(t^\prime)$. 
As usual $A_\mu (x) \equiv A_\mu^a (x) \lambda_a/2$, 
${\rm Tr}$ means the trace over colour indices,  ${\rm P}$
prescribes the ordering of the colour matrices according 
to the direction fixed on the loop and $S_{\rm YM}(A)$ is the Yang Mills 
action including a gauge fixing term. 

$\Delta V_{\rm a}$ is a gauge invariant quantity. 

The field-strength expectation values in (\ref{vab}), (\ref{vae}) can be 
expressed as functional derivatives of $\log \langle W(\Gamma) \rangle$ 
with respect to the path $\Gamma$ \cite{BCP}. 
In fact the change induced in $\langle W(\Gamma)\rangle$ by letting 
$z_1^\mu (t) \rightarrow  z_1^\mu (t) + \delta  z_1^\mu (t)$ where 
$\delta  z_1^\mu  (-T/2) = \delta  z_1^\mu (T/2) = 0$ is 
\begin{equation}
\langle\!\langle F_{\mu\nu}(z_1) \rangle\!\rangle = 
{\delta i \log \langle W(\Gamma) \rangle \over e\,\delta S^{\mu\nu} (z_1)} 
\, ,\label{var1}
\end{equation}

\begin{equation} 
\delta S^{\mu\nu} (z_1) =  
(dz_1^\mu \delta z_1^\nu - dz_1^\nu \delta z_1^\mu) \, , 
\end{equation}
and, varying the first quark line again, we have 
\begin{equation}
\langle\!\langle F_{\mu\nu}(z_1) 
F_{\lambda\rho}(z_1^\prime) \rangle\!\rangle 
- \langle\!\langle F_{\mu\nu}(z_1) \rangle\!\rangle 
\langle\!\langle F_{\lambda\rho}(z_1^\prime) \rangle\!\rangle = 
- {\delta^2 \log \langle W(\Gamma) \rangle \over 
e^2 \delta S^{\mu\nu} (z_1) \delta S^{\lambda\rho} (z_1^\prime) } \, .
\label{var12}
\end{equation}
Hence, the evaluation of (\ref{vab}) and (\ref{vae}) follows 
straightforwardly from the assumptions on the Wilson loop behaviour. 

In the definition of $\Delta V_{\rm a}$ (eqs (\ref{vab}), (\ref{vae})) 
the quarks are  static, i. e. $\zv_1$ and $\zv_2$ are fixed 
in time. Accordingly 
the lattice calculations have been done inserting field-strength tensors 
in a rectangular Wilson loop. 

In the static limit we note the following identities involving 
the static potential \cite{BMP}:
\begin{eqnarray}
\int_{-T/2}^{T/2} dt\, {\bf \nabla} V_0 &=& 
e \int_{-T/2}^{T/2} dt\, \langle\!\langle {\bf E}(z_1) \rangle\!\rangle 
\, ,\label{id1sta}\\
\int_{-T/2}^{T/2} dt\, \Delta V_0 &=& 
- i\, e^2 \int_{-T/2}^{T/2} dt\, \int_{-T/2}^{T/2} dt^\prime  
\langle\!\langle {\bf E}(z_1) \cdot {\bf E}(z_1^\prime) \rangle\!\rangle 
- \langle\!\langle {\bf E}(z_1) \rangle\!\rangle \cdot 
\langle\!\langle {\bf E}(z_1^\prime) \rangle\!\rangle  \nonumber\\
&~&\quad + e \int_{-T/2}^{T/2} dt\, 
 \langle\!\langle {\bf D} \cdot {\bf E}(z_1)\rangle\!\rangle
\, , \label{id2sta}
\end{eqnarray}
where $D_\mu$ is the covariant derivative. 
Therefore, we have the following alternative way to express 
$\Delta V_{\rm a}^B$ and $\Delta V_{\rm a}^E$: 
\begin{eqnarray}
\Delta V_{\rm a}^B &=& 
\lim_{T\to\infty} {i\over T} \int_{-T/2}^{T/2} dt 
\int_{-T/2}^{T/2}  dt^\prime   
{1\over 4}\epsilon_{ilm}\epsilon_{irs} 
{\delta^2 \log \langle W(\Gamma) \rangle \over \delta S_{lm}(z_1) 
\delta S_{rs}(z_1^\prime)} \, , \label{vab2}\\
\Delta V_{\rm a}^E &=&  - 
\lim_{T\to\infty} {i\over T} \int_{-T/2}^{T/2} dt 
\int_{-T/2}^{T/2}  dt^\prime   
{\delta^2 \log \langle W(\Gamma) \rangle \over \delta S_{0i}(z_1) 
\delta S_{0i}(z_1^\prime)} \, , \nonumber\\
&=& -\int_{-T/2}^{T/2} dt\, \Delta V_0 +  e \int_{-T/2}^{T/2} dt\,  
 \langle\!\langle {\bf D} \cdot {\bf E}(z_1)\rangle\!\rangle \, .
\label{vae2}
\end{eqnarray}
The last term in (\ref{vae2}) is the Darwin potential \cite{BCP,DQCDmal}.

\section{\bf Electric-magnetic duality}

We assume that for large loops the Wilson loop $W$ is the same as the Wilson 
loop $W_{\rm eff}$ determined by an effective dual theory which is weakly 
coupled at long distances \cite{DQCDmal}:
\begin{equation}
\langle W(\Gamma) \rangle = \langle W_{\rm eff}(\Gamma)\rangle 
\label{dual}
\end{equation}
where 
\begin{equation}
\langle W_{\rm eff} (\Gamma) \rangle= 
{\displaystyle{\int {\cal D} {\cal C} \,  {\cal D} {\cal B} \,
\exp \left[ i \int dx  {\cal L}_{\rm eff} (G_{\mu\nu}^{\rm S}) 
+ {\cal L}_{\rm GF} \right] }
\over \displaystyle{
\int {\cal D} {\cal C}\, {\cal D} {\cal B} \,  
\exp \left[ i \int dx  {\cal L}_{\rm eff} (G_{\mu\nu}^{\rm S}=0) 
+ {\cal L}_{\rm GF} \right]  }} \, .
\label{Weff}
\end{equation}
${\cal L}_{\rm eff}$ is the effective Lagrangian 
describing the interactions of dual potentials ${\cal C}_\mu$
 and monopole fields ${\cal B}_i$; the dual potentials interact 
with quarks via the Dirac string tensor $G_{\mu\nu}^{\rm S}$: 
\begin{equation}
G_{\mu\nu}^{\rm S}(x)  = 
e \,\epsilon_{\mu\nu\alpha\beta} \int ds \int d\tau\, 
{\partial y^\alpha \over \partial s} {\partial y^\beta\over \partial \tau}
\delta(x-y(s,\tau)) \, ,
\label{dft}
\end{equation}
where $y(s,\tau)$ is a world sheet with boundary $\Gamma$ swept out by the
Dirac string.  The colour electric 
field $D$ and the color magnetic field $H$ are the components of the 
dual field--strength tensor 
 ${\cal G}_{\mu\nu} = \partial_\mu {\cal C}_\nu 
- \partial_\nu {\cal C}_\mu + G^{\rm S}_{\mu\nu}$: 
 $D_i = {\cal G}_{0i}$ and $H_i = \hat{{\cal G}}_{0i}$.
The Duality assumption (\ref{dual}) yields expressions for the 
Yang Mills expectation values (\ref{var1}) and (\ref{var12})
in terms of expectation values  of the corresponding operators 
 in the dual theory \cite{DQCDmal}
\begin{eqnarray}
\langle\!\langle {\bf B}(z_1) \rangle\!\rangle &=&
{4\over 3}\langle\!\langle {\bf H}(z_1) \rangle\!\rangle_{\rm eff}   \, , 
\label{Heff}\\
\langle\!\langle {\bf E}(z_1) \rangle\!\rangle &=&
{4\over 3}\langle\!\langle {\bf D}(z_1) \rangle\!\rangle_{\rm eff}  \, ,
\label{Deff}\\
\langle\!\langle {\bf B}(z_1) \cdot {\bf B}(z_1^\prime) \rangle\!\rangle 
- \langle\!\langle {\bf B}(z_1) \rangle\!\rangle
\cdot \langle\!\langle {\bf B}(z_1^\prime) \rangle\!\rangle &=&
-{4\over 3}i\, {\delta \langle\!\langle H_j(z_1) \rangle\!\rangle_{\rm eff} 
\over \delta H^{\rm S}_j(z_1^\prime)} \, ,
\label{HHeff}\\
\langle\!\langle {\bf E}(z_1) \cdot {\bf E}(z_1^\prime) \rangle\!\rangle 
- \langle\!\langle {\bf E}(z_1) \rangle\!\rangle
\cdot \langle\!\langle {\bf E}(z_1^\prime) \rangle\!\rangle &=&
-{4\over 3}i\, {\delta \langle\!\langle D_j(z_1) \rangle\!\rangle_{\rm eff} 
\over \delta D^{\rm S}_j(z_1^\prime)} \, ,
\label{DDeff}
\end{eqnarray}
where $D^{\rm S}_i=G^{\rm S}_{0i}$, $H^{\rm S}_i=\hat{G}^{\rm S}_{0i}$. 
In the classical approximation to the dual  theory 
(justified because the theory is weakly 
coupled at large distances) all quantities on the right-hand side of 
equations (\ref{Weff}), (\ref{Heff})-(\ref{DDeff}) can be expressed in 
terms  of the solutions of the classical equations of motion, in particular 
\begin{eqnarray}
i\, \log \langle W(\Gamma) \rangle_{\rm eff} &=& 
- \int dx \, {\cal L}_{\rm eff} (G^{\rm S}_{\mu\nu}) \, ,
\label{claw}\\
\langle\!\langle {\bf D}(x) \rangle\!\rangle_{\rm eff} &=& 
{\bf D}({\bf x})  \, ,
\label{clad}\\
\langle\!\langle {\bf H}(x) \rangle\!\rangle_{\rm eff} &=& 
{\bf H}({\bf x})  \, .
\label{clah}
\end{eqnarray}

The explicit evaluation of (\ref{HHeff}) in the classical configuration 
gives \cite{DQCDmal}: 
\begin{equation}
\Delta V_{\rm a}^B = - {4\over 3} e^2 \left. 
\nabla_{\xv} \cdot \nabla_{\xv^\prime} 
G^{\rm NP}(\xv, \xv^\prime)
\right|_{\xv=\xv^\prime=\zv_1}\, .  \label{vabdqcd}
\end{equation}
where the function $G^{\rm NP}$ satisfies the equation 
\begin{equation}
\left(-\Delta_\xv + 6\, g^2 {\cal B}^2(\xv)\right) G^{\rm NP}(\xv,\xv^\prime) 
= - {6\, g^2 {\cal B}^2(\xv)\over 4\pi |\xv - \xv^\prime|} \, ,
\label{eqG}
\end{equation}
with $g = 2\pi/e$ and ${\cal B}$  is the static monopole field.

Now we use equation (\ref{eqG}) to obtain the behaviour 
of $G^{\rm NP}$ near the quark sources. Let us introduce the vector
\begin{equation}
{\bf K}(\xv) \equiv \left. \nabla_{\xv^\prime}
G^{\rm NP}(\xv,\xv^\prime) \right|_{\xv^\prime = \zv_1} \, . 
\end{equation}
In terms of cylindrical coordinate with origin in $\zv_1$, 
we have $\xv - \zv_1 = z\, {\bf e}_z + \rho\, {\bf e}_\rho$, where 
the unit vector ${\bf e}_z$ is directed along $\rv = \zv_1-\zv_2$ 
and the unit vector ${\bf e}_\rho$ is orthogonal to ${\bf e}_z$ 
in the plane of $\xv$ and $\rv$.  
By symmetry 
$$
{\bf K} = K_z {\bf e}_z + K_\rho {\bf e}_\rho \, .
$$ 
As a consequence, (\ref{eqG}) can be splitted into two equations, one for 
each component: 
\begin{eqnarray}
\left( -\Delta + 6\, g^2 {\cal B}^2(\xv)\right)K_z &=&
- {6\,g^2 z\, {\cal B}^2(\xv) \over 4\pi |\xv - \zv_1|^3} \, ,\label{Hz} \\
\left( -\tilde{\Delta} + 6\, g^2 {\cal B}^2(\xv)\right)K_\rho &=&
- {6\,g^2\rho\, {\cal B}^2(\xv) \over 4\pi |\xv - \zv_1|^3} \, ,  \label{Hr}
\end{eqnarray}
where, now, 
$\Delta = {\displaystyle {1\over \rho}{\partial \over \partial \rho}
+ {\partial^2\over \partial \rho^2} 
+ {1\over \rho^2}{\partial^2 \over \partial \phi^2} 
+ {\partial^2 \over \partial z^2}}$ and 
$\tilde{\Delta} \equiv \Delta - {\displaystyle{1\over \rho^2}}$.  
Let $\theta$ be the angle between $\rv$ and $\xv - \zv_1$, where 
 for $\theta = \pi$ the vector $\xv - \zv_1$ is along 
$\rv$, so that  $z = |\xv - \zv_1| \cos \theta$, 
$\rho = |\xv - \zv_1| \sin\theta$. On the string, ${\cal B}$ must vanish, 
and its behaviour near quark 1 is given by \cite{statspin} 
\begin{equation}
{\cal B} = b \sqrt{1+\cos\theta} |\xv - \zv_1|^{(\sqrt{3} - 1)/2} \, ,
\label{B}
\end{equation}
where $b$ is a function weakly dependent on $r$ (the quark-antiquark 
distance). In Table \ref{tableb} we list values of $b$ obtained by numerical 
integration of the static equations for ${\cal B}$ for varying separations 
$r$. Putting (\ref{B}) in (\ref{Hz})-(\ref{Hr}) and keeping only terms 
which are singular near the quark 1 (for $\xv\to \zv_1$), we obtain 
\begin{eqnarray}
-\Delta K_z &=& - {3\,g^2 b^2\over 2\pi}\cos\theta\,
(1 + \cos\theta) |\xv - \zv_1|^{\sqrt{3} - 3} \, ,\label{Hz1} \\
-\tilde{\Delta} K_\rho &=& - {3\,g^2 b^2\over 2\pi}\sin\theta\,
(1 + \cos\theta) |\xv - \zv_1|^{\sqrt{3} - 3} \, .\label{Hr1}
\end{eqnarray}
The solutions of these equations are \cite{statspin}
\begin{eqnarray}
K_z &=& - {3\,g^2 b^2\over 2\pi}\left( - {1\over 3} 
- {\cos\theta\over 1 - \sqrt{3}} + {\cos^2\theta\over 3 + \sqrt{3}}\right)
|\xv - \zv_1|^{\sqrt{3} - 1} \, ,\label{Hz2} \\
K_\rho &=& {3\,g^2 b^2\over 2\pi}\left( {1\over 1 - \sqrt{3}} 
- {\cos\theta\over 3 + \sqrt{3}}\right)\sin\theta 
|\xv - \zv_1|^{\sqrt{3} - 1} \, .\label{Hr2}
\end{eqnarray}
Finally 
\begin{eqnarray}
\Delta V_{\rm a}^B &=& - {4\over 3} e^2 
\left. \nabla_{\xv} \cdot \nabla_{\xv^\prime} 
G^{\rm NP}(\xv, \xv^\prime) \right|_{\xv=\xv^\prime=\zv_1} 
= {4\over 3} e^2 \nabla_{\xv} \cdot 
\left. {\bf K}(\xv) \right|_{\xv=\zv_1} \nonumber\\
&=& - 8\, \pi b^2 (\sqrt{3} +2) 
|\xv - \zv_1|^{\sqrt{3} - 2}\bigg|_{\xv = \zv_1} 
\, .\label{vabdres} 
\end{eqnarray}
The result (\ref{vabdres}) is divergent. However the dual theory has a natural 
ultraviolet cut-off which is the mass $M \approx 0.640$ GeV of the 
dual gluon, and we can put 
\begin{equation}
|\xv - \zv_1|^{\sqrt{3} - 2}\bigg|_{\xv = \zv_1} = M^{2 - \sqrt{3}} \, .
\end{equation}
From this prescription and the data of Table \ref{tableb}, 
we can derive the values of $\Delta V_{\rm a}^B$ at different distances 
as given in Table \ref{tablevab}. We observe that in the range of interest 
(0.2  fm $< r <$ 1 fm) the magnetic contribution to 
$\Delta V_{\rm a}$ can be considered almost a constant 
($\approx -$0.3 GeV$^3$).

In order to evaluate $\Delta V_{\rm a}^E$ in the dual theory
we use the following analogue of eq. (\ref{id2sta}) in the dual theory:
 \begin{eqnarray}
 & & \lim_{T\to \infty} {1\over T} \int_{-{T/2}}^{T/2} dt  
\int_{-{T/2}}^{T/2} dt^\prime
{4\over 3} e^2  {\delta \langle \langle D_j(z_1) \rangle\rangle_{\rm eff}
\over \delta D_j^{\rm S}(z_1^\prime)}=\nonumber\\
& & \quad \quad 
-\int_{-T/2}^{T/2} dt\, \Delta V_0 + {4\over 3} e \int_{-T/2}^{T/2} dt\,  
\nabla \cdot \langle\!\langle {\bf D}(z_1)\rangle\!\rangle_{\rm eff} \, . 
\label{vae2dual}
\end{eqnarray}
Comparing (\ref{vae2dual}) with (\ref{vae2}) we obtain
\begin{equation}
\int_{-T/2}^{T/2} dt\, 
 \langle\!\langle {\bf D} \cdot 
{\bf E}(z_1)\rangle\!\rangle = 
{4\over 3}\int_{-T/2}^{T/2} dt\,  
\nabla \cdot \langle\!\langle {\bf D}(z_1)\rangle\!\rangle_{\rm eff} \, .
\label{vae2ddual}
\end{equation}
Eq. (\ref{vae2ddual}) expresses the Darwin potential in terms of 
quantities calculated  in the dual theory and has a simple interpretation.
The left hand side  is the expectation value of the electric current 
operator in Yang Mills theory and the right hand side is the corresponding 
quantity in the dual theory (note that (\ref{vae2ddual}) and (\ref{Deff}) 
are independent results valid only on the quark trajectory). 
Since in the dual theory the colour electric field ${\bf D}$ is 
given by ${\bf D}= \nabla \times {\bf C} + {\bf D}^S$ the dual 
potential does not contribute to divergence of ${\bf D}$ and we have 
\begin{equation}
\nabla \cdot \langle\!\langle {\bf D}(z_1)\rangle\!\rangle_{\rm eff} 
= \nabla \cdot {\bf D}(z_1) = e \,\delta(\rv) \, ,
\label{divD}
\end{equation}
where $\rv = \zv_1 - \zv_2$. 
The dynamics of the dual theory  fixes  $\nabla \times {\bf D}$, that is,
the monopole
 current which, because of (\ref{vae2ddual}), does
not contribute to the Darwin term. Thus the monopole degrees of freedom, 
which in the classical approximation to the dual theory produce the dual 
Meissner effect and confinement, contribute to $\Delta V_{\rm a}$ only via 
the first term on the right hand side of eq.(\ref{vae2dual}). 
Thus comparison of $\Delta V_{\rm a}$ predictions with lattice data provides 
a crucial test of the dual description of long distance Yang Mills theory.

Putting (\ref{divD}) in (\ref{vae2dual}) we obtain
\begin{equation}
\Delta V_{\rm a}^E = -\Delta V_0 + {16\pi \over 3} \as \, \delta(\rv) 
= - \Delta \left( V_0 + {4\over 3}{\as\over r} \right) \, , 
\label{v0np}
\end{equation}
where $\as = e^2/4\pi$. 
The quantity $G^{\rm NP}$ determining $\Delta V_{\rm a}^B$ (\ref{vabdres}) 
is just the nonperturbative part of the monopole potential and so 
(\ref{vabdres}) is the natural magnetic counterpart of (\ref{v0np}). 
Thus both the electric and magnetic parts of $\Delta V_{\rm a}$ directly 
measure the characteristic feature of the dual picture. 
The static potential $V_0$ obtained by solving numerically the 
equations of motion can be represented by the following analytic 
parametrization [11]:
\begin{equation}
V_0(r) = -{4\over 3}{\as\over r} e^{-0.511 \, \sqrt{\sigma/\as} \, r}
+ \sigma r - 0.646\,\sqrt{\sigma\as} \, ,
\label{fit}
\end{equation}
where $\sigma \approx 0.2$ GeV$^2$ is the string tension. 
Substituting (\ref{fit}) in (\ref{v0np}) we obtain 
\begin{equation}
\Delta V_{\rm a}^E = 
{4\over 3}\,(0.511)^2{\sigma\over r} e^{-0.511 \, \sqrt{\sigma/\as}\, r}
- 2\, {\sigma\over r} \, .
\label{vaefinal}
\end{equation}
In Table \ref{tablevae} we list some values of $\Delta V_{\rm a}^E$  for 
different quark-antiquark distances. Notice that the dependence on $r$ 
is strong. In particular the short-range behaviour ($r\to 0$) is like 
$1/r$. In Table \ref{tableva} we sum up the magnetic and electric 
contributions and give the complete $\Delta V_{\rm a}$ behaviour. 
In the long-range region ($r\to\infty$) $\Delta V_{\rm a}$ has only magnetic 
contributions with an asymptotic value of 0.2 $\sim$ 0.3 GeV$^3$. 
In the short-range region $\Delta V_{\rm a}$ has a behaviour like $1/r$. 

\section{\bf Stochastic vacuum model}

In the stochastic vacuum model, one assumes that, in Euclidean space,  
the Wilson loop has the following behaviour \cite{DoSi}:
\begin{eqnarray}
&~& \log\langle W(\Gamma) \rangle = 
\nonumber\\
&~&\quad 
-{1\over 2} \int_S dS_{\mu\nu}(u) \int_S dS_{\lambda\rho}(v)  
(\delta_{\mu\lambda}\delta_{\nu\rho} - \delta_{\mu\rho}\delta_{\nu\lambda})
\left( D(h^2)+D_1(h^2) \right)
\nonumber\\
&~&\quad 
+ \left(h_\mu h_\lambda \delta_{\nu\rho} - h_\mu h_\rho \delta_{\nu\lambda} 
+ h_\nu h_\rho \delta_{\mu\lambda} - h_\nu h_\lambda \delta_{\mu\rho}\right) 
{\partial \over \partial h^2} D_1(h^2) \>, 
\label{svm}
\end{eqnarray}
where $h \equiv u-v$. The functions $D$ and $D_1$, called correlator 
functions, can be interpreted as vacuum form factors. The perturbative 
contributions (contained in $D_1$) can be evaluated by the usual $e^2$ 
expansion of the Wilson loop, while the nonperturbative contributions, 
dominant in the long-range behaviour come only from lattice calculations 
\cite{Dig}. The integration is performed over the minimal area surface 
$S$ with contour $\Gamma$. Up to order $1/m^2$ the minimal  surface can be 
identified exactly with the surface spanned by the straight-line 
joining the point $(t,{\bf z}_1(t))$ on the quark line 1 with the point 
$(t,{\bf z}_2(t))$ on the quark line 2 of $\Gamma$. Therefore, for 
our purposes, the generic point $u^\mu(t,s)$ on the surface $S$ 
will be given by:
\begin{equation}
u^0 = t \qquad\qquad\qquad {\bf u} = s\, {\bf z}_1(t) + (1-s)\,{\bf z}_2(t) 
\label{straight}
\end{equation}
and the surface element $dS_{\mu\nu}(u)$ by
\begin{eqnarray}
dS_{4j}(u)     &=& dt~ds~ r_j(t)       \>,
\label{surf1}\\
dS_{ij}(u)     &=& dt\,ds\, \left( s\,{\dot z}_{1i}(t) 
+ (1-s)\,{\dot z}_{2i}(t) \right)\,r_j(t) \>, 
\label{surf2}
\end{eqnarray}
where $t\in [-T/2, T/2]$, $s\in [0,1]$ and ${\bf r} = {\zv}_1 - {\zv}_2$. 

From equation (\ref{var12}) and (\ref{svm}), expanding 
\begin{equation}
z_1(t^\prime) = z_1(t) + (t-t^\prime)\, \dot{z_1}(t) + \cdots \, , 
\end{equation}
we have in the static limit (i. e. neglecting the velocity dependent terms) 
\begin{equation}
{\delta \log \langle W(\Gamma)\rangle \over 
\delta S_{lm}(z_1)\, \delta S_{rs}(z_1^\prime)} = 
-(\delta_{lr}\delta_{ms} - \delta_{ls}\delta_{mr}) 
\left( D(t^2) + D_1(t^2) \right) \, .
\label{BBsvm}
\end{equation}
Putting this expression in (\ref{vab2}) we obtain 
\begin{equation}
\Delta V_{\rm a}^B = -6 \int_0^\infty dt \,\left( D(t^2) + D_1(t^2)\right) \, .
\label{vabres}
\end{equation}

In an analogous way, since in the static limit 
\begin{equation}
{\delta \log \langle W(\Gamma)\rangle \over 
\delta S_{0i}(z_1)\, \delta S_{0i}(z_1^\prime)} = 
-3 \left( D(t^2) + D_1(t^2) + t^2 {d \over dt^2} D_1(t^2) \right) \, ,
\label{EEsvm}
\end{equation}
we obtain,
from the first equality in equation (\ref{vae2}) 
\begin{equation}
\Delta V_{\rm a}^E = -6 \int_0^\infty dt \, 
\left( D(t^2) + {1\over 2}D_1(t^2)\right) \, , 
\label{vaeres}
\end{equation}
where a partial integration was performed. 

In the usual long-range parameterization, suggested by the direct  
lattice evaluation of the correlator functions $D$ and $D_1$ \cite{BV,Dig}, 
\begin{eqnarray}
D(x^2) &=&d\, e^{ -\delta |x|}\, , 
\qquad\qquad \delta \approx 1\>{\rm GeV} \, ,
\qquad\quad d \approx 0.073 \> {\rm GeV}^4 \, , \label{Dpar}\\
D_1(x^2) &=&d_1\, e^{ -\delta_1 |x|}\, ,  
\qquad\quad \delta_1 \approx 1\>{\rm GeV} \, , 
\qquad\quad d_1 \approx 0.0254 \> {\rm GeV}^4 \, ,\label{D1par}
\end{eqnarray}
the electric and magnetic contribution to $\Delta V_{\rm a}$ are 
\begin{eqnarray}
\Delta V_{\rm a}^B &=& -6{d\over \delta} - 6 {d_1\over \delta_1} 
\approx -0.59 \> {\rm GeV}^3\, , \label{vabpar}\\
\Delta V_{\rm a}^E &=& -6{d\over \delta} - 3 {d_1\over \delta_1} 
\approx -0.51 \> {\rm GeV}^3\, , \label{vaepar}
\end{eqnarray}
and then 
\begin{equation}
\Delta V_{\rm a} = \Delta V_{\rm a}^B + \Delta V_{\rm a}^E \approx 
-1.1 \> {\rm GeV}^3 \>. \label{vapar}
\end{equation}

In contrast with the dual theory the long range  contributions 
to $\Delta V_{\rm a}$ in the stochastic vacuum model are self energy 
corrections independent of $r$.
For this reason these terms were neglected in \cite{BV} and 
did not appear in the minimal area law evaluation of \cite{BMP}. 
We expect that a $r$ dependence will arise in the short range 
region, taking into account higher order perturbative contributions. 
Since the $r$ dependence is very weak also in the dual theory for the 
magnetic contribution $\Delta V_{\rm a}^B$, the difference between 
the two models is in the electric contributions $\Delta V_{\rm a}^E$ 
and is mainly due to the Darwin term. In fact, in the static limit 
the dual theory predicts 
\begin{equation}
e \int_{-T/2}^{T/2} dt\, 
\langle\!\langle {\bf D}\cdot {\bf E}(z_1)\rangle\!\rangle  
= {4\over 3} e^2 \int_{-T/2}^{T/2} dt\,\delta(\rv) \, .
\label{divDdual}
\end{equation}
Eq.(\ref{divDdual}) reflects the fact that in the dual theory the color
electric field is generated by the monopole current which does not
contribute to the right hand side of eq.(\ref{vae2ddual}).
This contrasts with the corresponding result of the stochastic 
vacuum model:
\begin{equation}
e \int_{-T/2}^{T/2} dt\,
 \langle\!\langle {\bf D} \cdot {\bf E}(z_1)\rangle\!\rangle  
= \int_{-T/2}^{T/2} dt\, \nabla V_0 + \hbox{self-energy contributions.} 
\label{divDsvm}
\end{equation}

\section{\bf Conclusion}

The dual theory predictions of $\Delta V_{\rm a}$ are in very good agreement 
with the lattice data. The asymptotic value for large distances 
is negative and very well inside the error bars of the lattice prediction. 
Surprisingly also the short range prediction, which in principle could  
be beyond the validity range of the dual theory, reproduces the $1/r$ 
behaviour of the lattice fit. Also the long range result of the 
stochastic vacuum model is negative and compatible with the lattice data, 
but of self-energy type. A $r$ dependence arises from the perturbative 
contributions only at the next-to-leading order. 
However the main contribution to the ${1\over r}$ behavior of 
$V_{\rm a}^E$ in the dual theory comes from the second term on the
right hand side of eq. (\ref{vaefinal}), which is clearly nonperturbative.
Furthermore, as is seen from Table 2, $V_{\rm a}^B$ also increases with
decreasing $r$.
After this calculation 
the main difference in the two models seems to be clear. 
In the dual theory only the quark charges contribute in the static limit 
to the Darwin potential, while in the stochastic vacuum model 
the Darwin potential is related to the full static potential. 
This means that a truly accurate lattice determination of
the field correlator (which is possible for the magnetic part
of the correlator and it is already in progress \cite{balic})
is of great interest and could provide significant information
concerning the nonperturbative quark interaction.

\vfill\eject

\begin{table}[htb]{}
\caption{$b$ vs $r$, the $q \bar{q}$ distance. The table shows 
a slight dependence.}
\label{tableb}
\vspace{0.5in}
\begin{tabular}{|l|l|}
\hline
$r$ (fm) \qquad\qquad &$b^2/M^{\sqrt{3}-1}$ (GeV$^2$) \\ 
\hline
0.175  &\qquad 0.007 \\ 
0.26   &\qquad 0.006 \\ 
0.35   &\qquad 0.0055 \\ 
0.7    &\qquad 0.0048 \\ 
1.4    &\qquad 0.0045 \\
\hline
\end{tabular}
\end{table}

\vskip 5 truecm 

\begin{table}[hbt]{}
\caption{$\Delta V_{\rm a}^B$ at the ultraviolet cut-off 
$M = 0.640$ Gev vs $r$, the $q \bar{q}$ distance
($\as=0.37$ and $\sigma=0.2 \, {\rm Gev}^2$).}
\label{tablevab}
\vspace{0.5in}
\begin{tabular}{|l|l|}
\hline
$r$ (fm) \qquad\qquad &$\Delta V_{\rm a}^B$ (GeV$^3$) \\ 
\hline
0.175  &\quad $-$ 0.407 \\ 
0.26   &\quad $-$ 0.350\\ 
0.35   &\quad $-$ 0.320 \\ 
0.7    &\quad $-$ 0.277 \\ 
1.4    &\quad $-$ 0.263 \\
\hline
\end{tabular}
\end{table}

\vfill\eject

\begin{table}[thb]{}
\caption{$\Delta V_{\rm a}^E$ vs $r$, the $q\bar{q}$ distance 
($\as=0.37$ and $\sigma=0.2 \,{\rm Gev}^2$).}
\label{tablevae}
\vspace{0.5in}
\begin{tabular}{|l|l|}
\hline
$r$ (fm) \qquad\qquad &$\Delta V_{\rm a}^E$ (GeV$^3$) \\ 
\hline
0.175  &\quad $-$ 0.402 \\ 
0.26   &\quad $-$ 0.275\\ 
0.35   &\quad $-$ 0.208 \\ 
0.7    &\quad $-$ 0.109 \\ 
1.4    &\quad $-$ 0.057 \\
\hline
\end{tabular}
\end{table}

\vskip 5 truecm 

\begin{table}[hbt]{}
\caption{$\Delta V_{\rm a} = \Delta V_{\rm a}^B + \Delta V_{\rm a}^E$ vs 
$r$, the $q\bar{q}$ distance ($\as=0.37$ and $\sigma=0.2 \, {\rm Gev}^2$).}
\label{tableva}
\vspace{0.5in}
\begin{tabular}{|l|l|}
\hline
$r$ (fm) \qquad\qquad &$\Delta V_{\rm a}$ (GeV$^3$) \\ 
\hline
0.175  &\quad $-$ 0.809 \\ 
0.26   &\quad $-$ 0.625\\ 
0.35   &\quad $-$ 0.528 \\ 
0.7    &\quad $-$ 0.386 \\ 
1.4    &\quad $-$ 0.320 \\
\hline
\end{tabular}
\end{table}

\end{document}